\documentclass{SCGE}
\usepackage[colorlinks,linkcolor=blue,citecolor=blue,urlcolor=blue]{hyperref}
\usepackage{ulem}

\let\citedash\relax
\makeatletter \providecommand{\citedash}{\hbox{-}\penalty\@m}
\makeatother

\begin{document}
\begin{picture}(0,0){\rm
\put(0,-20){\makebox[160truemm][l]{\bf {\sanhao\raisebox{2pt}{.}}
Article  {\sanhao\raisebox{1.5pt}{.}}}}}
\put(0,-34){\jiuwuhao {\textcolor[rgb]{0.5,0.5,0.5}{\sf 
}}}
\end{picture}

\def\bm{\boldsymbol}

\def\dl{\displaystyle}
\def\du{\end{document}}
\def\d{{\rm d}}
\def\e{{\rm e}}
\def\i{{\rm i}}

\Year{2016} %
\Month{January} %
\Vol{59} 
\No{1} 
\BeginPage{1} 
\AuthorMark{{\rm A. Author}, et al.}  
\DOI{} 
\ArtNo{000000}

\title[The First Pulsar of FAST]{The First Pulsar Discovered by FAST}

\author[1]{Lei Qian$^*$}{}
\author[1]{Zhichen Pan}{}
\author[1,2]{Di Li$^*$}{}
\author[1,3]{George Hobbs}{}
\author[1]{Weiwei Zhu}{}
\author[1]{Pei Wang}{}
\author[4]{Zhijie Liu}{}
\author[1]{Youling Yue}{}
\author[1]{\\Yan Zhu}{}
\author[1]{Hongfei Liu}{}
\author[1]{Dongjun Yu}{}
\author[1]{Jinghai Sun}{}
\author[1]{Peng Jiang}{}
\author[1]{Gaofeng Pan}{}
\author[1]{Hui Li}{}
\author[1]{Hengqian Gan}{}
\author[1]{\\Rui Yao}{}
\author[4]{Xiaoyao Xie}{}
\author[5]{Fernando Camilo}{}
\author[3]{Andrew Cameron}{}
\author[2]{Lei Zhang}{}
\author[2]{Shen Wang}{}
\author[]{FAST Project}{}

\address[{\rm1}]{CAS Key Laboratory of FAST, National Astronomical Observatories, Chinese Academy of Sciences,
             Beijing 100101, China; {\it lqian@nao.cas.cn; dili@nao.cas.cn};}
\address[{\rm2}]{School of  Astronomy and Space Science, University of Chinese Academy of Sciences,,Beijing100049, China. }
\address[{\rm3}]{Commonwealth Scientific and Industrial Research Organization, Australia Telescope National Facility, Sydney 2122, Australia.}
\address[{\rm4}]{Key Laboratory of Information and Computing Science, Guizhou Normal University, Guiyang 550001, China.}
\address[{\rm5}]{South African Radio Astronomy Observatory, Observatory 7925, South Africa.}

\maketitle \vspace{-3.5mm}{\footnotesize\begin{center} Received January 1, 2016; accepted January 1, 2016; published online January 1, 2016
\end{center}}\vspace*{-5mm}



\begin{center}
\CITA    
\end{center}

\textwidth=178truemm \textheight=236truemm

\wuhao\vspace*{1.5mm}

\begin{multicols}{2}

\renewcommand{\baselinestretch}{1.08} \baselineskip 12.2pt\parindent=10.8pt

\renewcommand{\thefootnote}

To assist with the commissioning \cite{yao2019} of the Five-hundred-meter Aperture Spherical radio Telescope (FAST), we performed a pulsar search, with the primary goal of developing and testing the pulsar data acquisition and processing pipelines. We tested and used three pipelines, two  (P1 and P2 hereafter) searched for the periodic signature of pulsars whereas the other one was used to search for bright single pulses (P3 hereafter). A pulsar candidate was discovered in the observation on the 22$^{nd}$ August, 2017, and later confirmed by the Parkes radio telescope on the 10$^{th}$ September, 2017. This pulsar, named PSR~J1900$-$0134, was the first pulsar discovered by FAST. The pulsar has a pulse period of 1.8\,s and a dispersion measure (DM) of 188\,pc\,cm$^{-3}$.


The search for radio pulsars is one of the main science pursuits for the Five-hundred-meter Aperture Spherical radio Telescope (FAST, \cite{nan}). With the growing population of pulsars, we are finding pulsars exhibiting interesting phenomena, such as pulsars whose emission switches off, those where the emission drifts in phase, and much more \cite{clifton,manchester2001}. Pulsars are also used to provide stringent tests of theories of gravity, to search for errors in terrestrial time standards and even to navigate spacecraft \cite{nicolas,deng2013}. Of course, there are still new pulsars and new phenomena waiting to be discovered (e.g., pulsar-black hole binary systems). Such pulsars will be found through more sensitive surveys that cover wide regions of the sky.

Commissioning observations with the FAST telescope, in order to prepare the telescope for a large-scale survey, have been ongoing since the 25$^{th}$ September, 2016. These commissioning observations have allowed us  to test the performance of the telescope and the data processing systems.

The majority of these commissioning observations have been carried out with a wide-band receiver system in a drift-scan observing mode where the receiver is fixed in the sky. The first data processing system that was commissioned provided high-time-resolution recordings of the incoming signal. Such data streams are ideal for radio pulsars searches. We have searched in all the recorded data for the signatures of pulsars.

These commissioning observations have been phenomenally successful. More than 60 pulsar candidates have been discovered and more then 50 of them were confirmed as new pulsars$^*$\footnote{*http://crafts.bao.ac.cn/pulsar}. In this {\it letter}, we present the first pulsar discovered by FAST.


The Commensal Radio Astronomy FAST Survey (CRAFTS$^{**}$\footnote{**http://crafts.bao.ac.cn}, \cite{li2018}) began in August 2017. Before June 2018, it was a drift-scan survey using a 270-1620\,MHz wide-band single-beam receiver. The system temperature (without the contribution from sky) was about 60-70 K, and the gain wass estimated to be 10.1 K/Jy during the observation.

The incoming data stream was processed by a digital backend system based on a Field Programmable Gate Array (FPGA) board (known as the ``ROACH2'' board) developed by the Collaboration for Astronomy Signal Processing and Electronic Research (CASPER)$^{\dag}$\footnote{{\dag}https://casper.berkeley.edu/}.
The entire band of the receiver was separated into two parts, which were processed separately and recorded in two separate PSRFITS-format \cite{hotan2004} files. The first file covered a frequency range of 270-1024\,MHz and the other covered 1024 to 1620\,MHz, both with the channel width of 0.25 MHz.

Drift scan observations were generally performed throughout the night (from 7pm to 8am). The data was split into individual files each 2\,GB in size. The data rate was 80\,MB/s in total, implying that for one hour there would be 288\,GB of data. We noted that a given source will drift through the beam in $\sim 10$\,s and $\sim$60\,s for the high- and low-frequency ends of the band respectively.

The resulting data files were first stored on the servers on the FAST site before transferred to the FAST early science data center at Guizhou Normal University through a dedicated 2\,Gb/s Ethernet connection.

The drift-scan observations that have been completed covered about 25\% of the FAST sky, whose declination range is -14.4$^\circ$ to 65.6$^\circ$.  The choice of the declination tracks was made through a population synthesis of the most likely areas of the sky to find new pulsars (in brief, this simulation suggested regions close to the Galactic plane, but outside of the Arecibo sky).


In order to carry out a search for the pulse periodicity, the raw data files were combined to form data files each with the length of 52.4288 seconds (this time corresponded to a pulsar drifting through the center of the beam at $\sim 300$\,MHz). In order to maximise sensitivity to pulsars drifting through the beam, the data files were formed every 26\,s to ensure overlapping data in adjacent files.

During the commissioning analysis the entire band was analysed. We found that the most sensitive band for searching for new pulsars was between 290 and 340\,MHz (corresponding to 200 channels over 50\,MHz bandwidth). This band was chosen based on the overall consideration of the larger beam size at lower frequencies, relatively little radio frequency interference (RFI), the limited computing power available and that pulsars were typically brighter at lower observing frequencies. The data were processed by a cluster equipped with 480 cores (forty E5-2680 v3 CPUs), with a pulsar search pipeline (P1) based on the PRESTO$^{\ddag}$\footnote{\ddag https://www.cv.nrao.edu/~sransom/presto/} software package\cite{ransom}. A separate cluster of desktop computers, provided by Guizhou Normal University, was used for searching pulsars in the data covering the wider band between 290 and 802\,MHz (512\,MHz bandwidth), also with a PRESTO-based pipeline (P2). These two pipelines used similar period and DM trials, but different frequency bands.

The details of these two PRESTO-based pipelines will be described in a later paper. In brief, we searched for pulsars with \textsc{rfifind} to remove RFIs and then used \textsc{accelsearch} to search for pulsars with dispersion measures up to $\sim 2000$\,cm$^{-3}$pc. With the clusters mentioned above, the real data processing time was much shorter than the observing time. Thus, in principle, we were able to perform the searches for periodic signals in real time.

The single pulse search pipeline (P3) was described by $^{\dag\ddag}$\footnote{$^{\dag\ddag}$W.W. Zhu et al. in preparation (2018)}. The pipeline used GPUs running on the cluster at the FAST site, processing data in the band of 270 to 526\,MHz (with a bandwidth of 256\,MHz). This pipeline was fast and the data can be processed in real time.

A drift scan survey during one night can last for 10 hours or more. The resulting data can provide more than 30000 candidates from one periodicity search and thousands of candidates from the single pulse search. We used three ranking methods. The first method was looking at every candidate by eye. The second method was using a tool based on the REAPER method \cite{faulkner2004} in which the candidates were plotted as a function of signal-to-noise radio (S/N) and period or dispersion measure. Those candidates with the unique period and relatively high SNR were highly possible to be real pulsars. The third method was using a machine-learning algorithm (PICS \cite{zhu}). Similar methods were used for ranking the single pulse candidates.

A highly ranked pulsar candidate (PSR~J1900$-$0134)
was identified with a rotation period of 1832.306\,ms and a dispersion measure of 188.3\,pc\,cm$^{-3}$ (the original name for the pulsar candidate was PSR~J1859$-$01). This candidate was found in the data taken on the 22$^{th}$ August 2017 with all the three pipelines, i.e. two periodic search pipelines (P1 and P2) and one single pulse search pipeline (P3). Provisional parameters for this pulsar (a detailed description of our follow-up timing observations will be reported elsewhere) were listed in Table~2. The right ascension and the declination were determined by the azimuthal angle of the telescope feed, the elevation and the time when the flux maximum occurs during the drift-scan. The accuracy in right-ascension in the original discovery observation was better than 1', while the accuracy in declination was better than 2'. However, the timing observations have enabled us to determine the position with more precision.

\begin{figure}[h]
\centering
\includegraphics[scale=0.2]{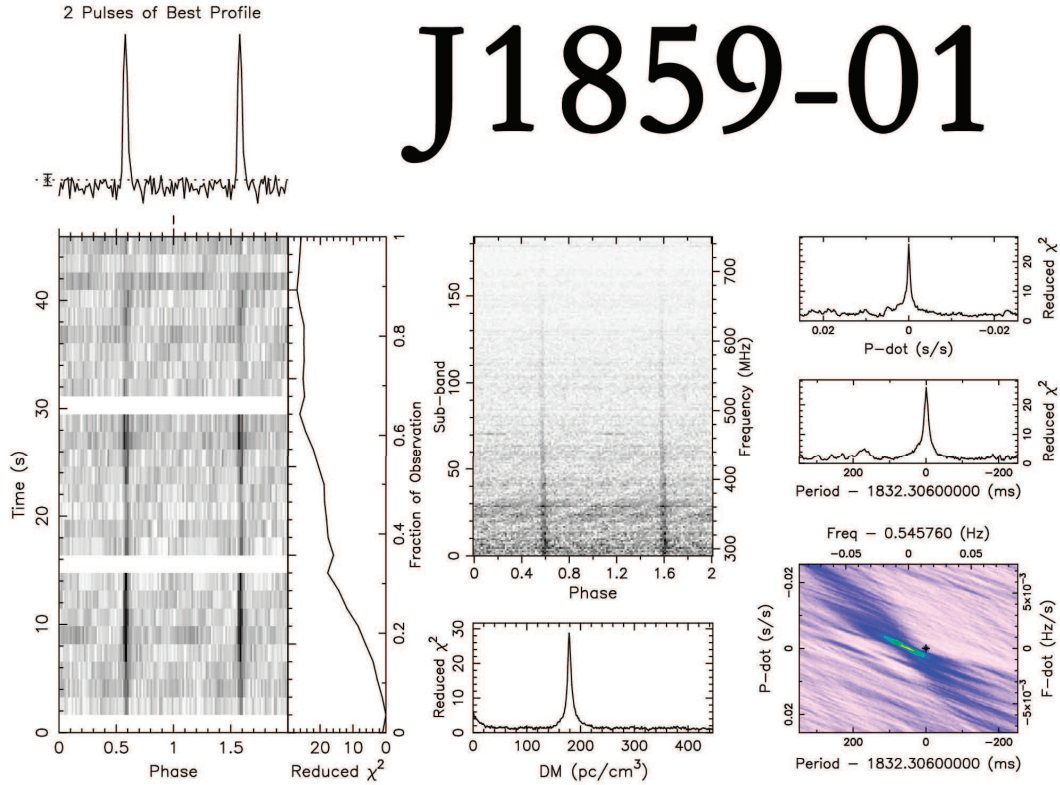}
\caption{The folded pulse profile (top left), intensity (gray-scale) of the pulse as a function of pulse phase and integration time (bottom left),  intensity as a function of pulse phase and observing frequency (top middle), a measure of the signal-to-noise of the pulse profile as a
function of dispersion measure (bottom middle) of PSR~J1900$-$0134 (note that as this is the discovery plot we indicate the pulsar's original name of PSR~J1859$-$01). The right-hand plots show small-scale searches to identify the pulse period and the period derivative that leads to the folded pulse profile with the highest signal-to-noise ratio.}
\label{J1859-01}
\end{figure}

In Figure~\ref{J1859-01} we show the discovery image of the pulsar by pipeline P2. This figure contains the folded pulse profile (top left), the intensity (gray-scale) of the pulse as a function of pulse phase and integration time (bottom left) and the intensity as a function of pulse phase and observing frequency (top middle). A statistic value ($\chi^2$, which can be treated in a similar fashion to the signal-to-noise ratio of the pulse profile) as a function of dispersion measure is given in the bottom/middle plot. The right-hand plots show small-scale searches to identify the pulse period and the period derivative that leads to the folded pulse profile with the highest signal-to-noise ratio.

As this was the first major discovery made using FAST we wished to confirm the discovery using an independent telescope. We therefore re-observed the pulsar with the Parkes telescope in Australia on the 10$^{th}$ September, 2017. We also note that a reprocessing of the Parkes Multibeam Pulsar Survey data had a pulsar candidate at this position in the sky with similar period and dispersion measure. The Parkes telescope confirmed the discovery and monitoring observations of the pulsar are ongoing and will be published elsewhere.


PSR~J1900$-$0134 was the first pulsar discovered by FAST. We have already confirmed a large number of other pulsars and the number of discoveries is expected to rapidly increase as we complete the commissioning of the telescope system and start our major all-sky survey with the recently installed 19-beam receiver. Many of the pulsars discovered have already been found to be intrinsically interesting and such pulsars will soon be described in papers currently being prepared. The enormous gain of the FAST telescope along with the short integration times and powerful data processing systems provides FAST with the capability of finding very extreme systems including sub-millisecond pulsars and close binary systems, potentially including pulsar-black-hole binaries.

\vspace*{2mm} \Acknowledgements{\bahao This work is supported by National Key R\&D Program of China (No. 2017YFA0402600, 2015CB857100), the Open Project Program of the Key Laboratory of FAST, NAOC, Chinese Academy of Sciences, the Strategic Priority Research Program of the Chinese Academy of Sciences Grant No. XDB23000000, and by the National Natural Science Foundation of China under grant No. 11690024, 11743002, 11873067, 11725313, and the CAS International Partnership Program No. 114A11KYSB20160008.
LQ is supported in part by the Youth Innovation Promotion Association of CAS (id.~2018075).
ZCP is supported by the National Natural Science Funds of China (Grant No. 11703047) and the CAS "Light of West China" Program.
YLY is supported by CAS "Light of West China" Program.
WWZ is supported by the Chinese Academy of Science Pioneer Hundred Talents Program. The Parkes radio telescope is part of the Australia Telescope National Facility which is funded by the Australian Government for operation as a National Facility managed by CSIRO.
}

\end{multicols}

\begin{table}
\centering
  \caption{FAST specifications.}
  \begin{tabular}{lc}
  \hline
  Specifications & Value \\
  \hline
  Declination range  & 14.4$^\circ$-65.6$^\circ$  \\
  System temperature $T_{\rm sys}$ (without the contribution from sky $T_{\rm sky}$) & 60-70 K \\
  Frequency range (ultra-wide band receiver) & 270 MHz-1620 MHz \\
  Number of polarization & 2\\
  sampling time & 200 $\mu$s\\
  \hline
  \label{FAST_info}
\end{tabular}
\end{table}


\begin{table}
\caption{Parameters of PSR~J1900$-$0134 from the timing observation of the Parkes telescope.}
\begin{center}
\begin{tabular}{lc}
\hline
Properties & Value \\
\hline
Right ascension, RA (J2000) & 19:00:26.02(7) \\
Declination, Dec (J2000) & $-$01:34:38(2) \\
Spin period, $P$ (s) & 1.8323313839(15)\\
Spin period derivative, $\dot{P}$ & $3.040(8)\times10^{-14}$\\
Timing epoch (MJD) & 58007 \\
Dispersion measure, DM ($\text{pc}\,\text{cm}^{-3}$) & 179(2) \\
Data span (MJD) & $58008-58395$ \\
Number of TOAs & 103\\
RMS residual ($\mu\text{s}$) & 4246 \\
\hline
\label{Pulsar_info}
\end{tabular}
\end{center}
\end{table}


\end{document}